\documentclass{ws-ijmpa}

\usepackage{amssymb}
\usepackage{amsmath}

\def\2{{1\over 2}}

\newcommand{\rf}[1]{(\ref{#1})}
\def\b{\bar}
\newcommand{\ud}{\mathrm{d}}

\newcommand{\p}{\partial}

\newcommand{\mA}{\mathbf{A} }
\newcommand{\mB}{\mathbf{B} }
\newcommand{\mC}{\mathbf{C} }
\newcommand{\mD}{\mathbf{D} }
\newcommand{\mW}{\mathbf{W} }
\newcommand{\mU}{\mathbf{U} }
\newcommand{\mY}{\mathbf{Y} }
\newcommand{\mF}{\mathcal{F} }
\newcommand{\mH}{\mathcal{H} }
\newcommand{\mQ}{\mathcal{Q} }
\newcommand{\m}{\mathfrak{m}}
\newcommand{\mg}{\mathfrak{g}}
\newcommand{\di}{\mathbf{div}}

\begin{document}
\markboth{A.M. Zeitlin}
{BV Yang-Mills Theory as a Homotopy Chern-Simons Theory via SFT}

%
\catchline{}{}{}{}{}
%

\title{BATALIN-VILKOVISKY YANG-MILLS THEORY\\ AS A HOMOTOPY CHERN-SIMONS THEORY VIA\\ 
STRING FIELD THEORY}

\author{ANTON M. ZEITLIN\footnote{http://math.yale.edu/$\sim$az84,  http://www.ipme.ru/zam.html}} 

\address{Department of Mathematics, Yale University,\\
442 Dunham Lab, 10 Hillhouse Ave,
New Haven, CT 06511, USA\\
anton.zeitlin@yale.edu}

\maketitle

\begin{history}
\received{30 July 2008}
\end{history}

\begin{abstract}
We show explicitly how Batalin-Vilkovisky Yang-Mills action emerges as a homotopy generalization 
of Chern-Simons theory from the algebraic constructions arising from string field theory. 

\end{abstract}

\ccode{PACS numbers: 11.25.Hf, 11.25.Sq, 11.30.Na}

\section{Introduction: Chern-Simons vs Yang-Mills}
Chern-Simons-like theories have played the important role in both quantum field theory and string theory for a long time. 
The interest in such theories started from the original 3d Chern-Simons theory\cite{schwarz}\cdash\cite{deser3} 
with the action functional
\begin{eqnarray}
S_{CS}=\int_{M^3} Tr\Big(\frac{1}{2}\mathbf{A}\wedge\ud\mA+\frac{1}{3!}[\mA\wedge\mA]\wedge\mA\Big),
\end{eqnarray}
which was one of the first considered topological field theories. Soon after that, the generalized Chern-Simons 
theories with the action 
\begin{eqnarray}\label{cs2}
S^{gen}_{CS}=\frac{1}{2}\langle \Psi, \mQ\Psi \rangle+\frac{1}{3!}\{\Psi,\Psi,\Psi\},
\end{eqnarray}
where $\mQ$ is some nilpotent operator, 
$\langle\cdot,\cdot\rangle$ is some pairing and $\{\cdot,\cdot,\cdot\}$ is some graded 
(anti)symmetric three-linear operation, drew a lot of attention, for example in relation to 
open string field theory.\cite{witten} 

The equations of motion for the theories of type \rf{cs2} are generalized Maurer-Cartan equations
\begin{eqnarray}
 \mQ\Psi+\frac{1}{2}[\Psi,\Psi]=0,
\end{eqnarray}
where $[\cdot,\cdot]$ is some graded (anti)symmetric bilinear operation, which in the case of the usual 
3d Chern-Simons theory is simply
the zero curvature equation. 

In Ref.~\refcite{ym}, motivated by the well-known fact from open SFT\cite{siegel}\cdash\cite{taylor1}, 
that the Maxwell equations 
are given by the linear equation
\begin{eqnarray}
Q\phi_{\mA}=0,\quad  \phi_{\mA}=(-ic_1A_{\mu}(x)a^{\mu}_{-1}-c_0\p^{\mu}A_{\mu}(x))|0\rangle,
\end{eqnarray}
where $Q$ is a BRST operator of open string,\cite{brst}\cdash\cite{furuuchi} 
we gave a homological meaning to general Yang-Mills equations. Namely, we considered the subcomplex $\mF$ of 
the BRST complex of open SFT 
and constructed the graded (w.r.t. the ghost number) operations 
$[\cdot,\cdot]_h$, $[\cdot,\cdot,\cdot]_h$ on $\mF_{\mg}=\mF\otimes \mg$ ($\mg$ is some reductive Lie algebra), 
which together with the BRST operator formed a 
homotopy Lie algebra. The Yang-Mills equations 
\begin{eqnarray}
\p_{\mu}F^{\mu\nu}+[A_{\mu},F^{\mu\nu}]=0
\end{eqnarray}
appeared to be the generalized Maurer-Cartan equations associated with this 
homotopy Lie algebra: 
\begin{eqnarray}
Q\phi_{\mA}+\frac{1}{2}[\phi_{\mA},\phi_{\mA}]_h+\frac{1}{3!}[\phi_{\mA},\phi_{\mA}, \phi_{\mA}]_h=0.
\end{eqnarray}
In this paper, we  write the action for the Yang-Mills theory and also its Batalin-Vilkovisky (BV) 
version,\cite{bv}\cdash\cite{bv2}
\begin{eqnarray}\label{bvym}
&&S^{BV}_{YM}=\int d^Dx\Big(\frac{1}{2}(F_{\mu\nu}(x)F^{\mu\nu}(x))_K+2(D_{\mu}\omega(x),A^{*\mu}(x))_K-\nonumber\\
&&([\omega(x),\omega(x)],\omega^*(x))_K\Big),
\end{eqnarray}
where $(\cdot, \cdot)_K$ is a canonical invariant form on $\mg$, in the form of what we call a $homotopy$ $Chern$-$Simons$ $action$:
\begin{eqnarray}\label{hcs}
S_{HCS}=-\frac{1}{2}\langle\Phi, Q\Phi\rangle-\frac{1}{3!}\{\Phi,\Phi,\Phi\}_h-\frac{1}{4!}\{\Phi,\Phi,\Phi,\Phi\}_h,
\end{eqnarray}
where $\langle\cdot,\cdot\rangle$ is an appropriately defined pairing, and $\{\cdot,...,\cdot\}_h=\langle [\cdot,...,\cdot]_h,\cdot\rangle$ 
are the graded antisymmetric n-linear 
operations (n=3,4) on $\mF_{\mg}$. In the language of open SFT, $\Phi\in \mH_{\mg}$, which 
leads to the BV Yang-Mills action, 
has the following form: 
\begin{eqnarray}\label{state}
&&\Phi(\omega, \omega^*, \mA, \mA^*)=\nonumber\\
&&(\omega(x)-ic_1A_{\mu}(x)a_{-1}^{\mu}-c_0\p_{\mu}A^{\mu}(x)+
ic_1c_0 A^*_{\mu}(x)a_{-1}^{\mu}-c_1c_0 c_{-1} \omega^*(x))|0\rangle,
\end{eqnarray}
where $\omega$ is the ghost field and $\omega^*, \mA^*$ are antifields of the ghost and gauge field. 
We note here that in  Refs.~\refcite{taylor} and \refcite{berkovits},
 the non-Abelian versions of Yang-Mills actions were 
obtained from the effective actions of canonical open SFT\cite{witten} and WZW-like super-SFT\cite{berksft} correspondingly. 
In Ref.~\refcite{feng}  the BV Yang-Mills action up to three-point terms was obtained from open SFT. 
It would be interesting to find out how our constructions are related to the algebraic structure of open SFT.

Continuing our comparison of Yang-Mills and Chern-Simons theories, we recall that the 
construction of BV quantization of the usual 3d Chern-Simons theory leads to the action (see e.g. Refs.~\refcite{schkonts} 
and  \refcite{krotovlosev}): 
\begin{eqnarray}
S^{BV}_{CS}=\int_{M^3} Tr\Big(\frac{1}{2}\Psi\ud \Psi+\frac{1}{3}\Psi^3\Big),
\end{eqnarray}
where
\begin{eqnarray}
\Psi=\omega+A_{\mu}dx^{\mu}+\frac{1}{4}\epsilon^{\mu\nu\rho}A_{\mu}^*dx^{\nu}\wedge dx^{\rho}+\frac{1}{24}\epsilon^{\mu\nu\rho}
\omega^*dx^{\mu}\wedge dx^{\nu}\wedge dx^{\rho}.\label{form}
\end{eqnarray}
One can see the evident similarity between the expressions \rf{state} and \rf{form}. 
 
Finally, we note that the action \rf{hcs} recalls the action for the closed SFT, constructed by 
Zwiebach,\cite{zwiebach} which in comparison with the action \rf{hcs} contains infinite number of terms. 

The outline of the paper is as follows. In Sec. 2, we first recall some constructions we considered in Ref.~ \refcite{ym}. 
We give the definition of the Yang-Mills chain complex $\mF_{\mg}$ and show that the BRST operator $Q$ can be reduced to the 
$sl(2,\mathbb{R})$ Chevalley operator. We also give a new realization of this complex which allows us to construct the graded 
(w.r.t the ghost number) symmetric pairing. Then, we define the two-linear and three-linear graded 
antisymmetric operations which satisfy, 
together with the operator $Q$, the homotopy Lie algebra\cite{stasheff} relations. To formulate the homotopy Chern-Simons action, 
we introduce the multilinear operations $\{\cdot, ..., \cdot\}_h$ which we have mentioned above. After that, we 
show that the pure Yang-Mills action can be reformulated in the form \rf{hcs}. In the last part of Sec 2, we demonstrate 
how we arrived 
at the multilinear operations generating the homotopy Lie algebra; namely, we show how it emerges from the OPE in the boundary CFT of 
open string on the upper half-plane.\cite{zeit3} 

The main result of Sec. 3 is the formulation of the BV Yang-Mills action \rf{bvym} in the form \rf{hcs}. To do this, first of 
all we consider a tensor product of the complex $\mF_{\mg}$ with some Grassmann algebra $A$ to introduce the degrees of freedom of 
fermion statistics. Then, it is reasonable to change the grading in the complex and consider the total ghost number (taking into account 
$\mathbb{Z}$-gradation in $A$) making the resulting complex $\mH_{\mg}=\mF_{\mg}\otimes A$  infinite. After that, we appropriately 
redefine all algebraic structures we constructed for $\mF_{\mg}$ in the case of $\mH_{\mg}$. At the end of Sec. 3, we find the 
correspondence 
between \rf{bvym} and \rf{hcs}. 

In Sec. 4, we give a quick review of the BV formalism\cite{bv}$^,$\cite{bv2}$^,$\cite{schbv}$^,$\cite{stasheff2}
 and consider the gauging of BV Yang-Mills 
action in the homotopy Chern-Simons form. The paper ends with final remarks.

It should be noted that the local version (when all the fields are constant) of 
$L_{\infty}$ algebras, corresponding to Yang-Mills theory, 
was considered in Refs.~\refcite{movshev} and \refcite{movshev2}. 

One should also note that since the original paper Ref.~ \refcite{schkonts}, where the abstract statement about
 the relation between the BV formalism 
and $L_{\infty}$ algebras was given, there has been a lack of consideration of explicit field theory examples (except for 
Refs.~\refcite{movshev} and \refcite{movshev2} the explicit 
examples were different versions of Chern-Simons theory). We hope that here we partly fill this gap.

In this paper, our main goal is to point out the relation between the structures coming from open SFT  
and the BV formalism of pure Yang-Mills theory. 
In a separate paper, Ref.~ \refcite{ext}, we shall consider the extension of our results 
to the case of scalar and fermion fields coupled with gauge theory.

\section*{Notation and Conventions} 
\underline{\bf BRST operator in open String theory.}
In the case of open string theory in dimension $D$, one has $D$ scalar fields $X^{\mu}(z)$ such that the 
mode expansion is
\begin{eqnarray}
&&X^{\mu}(z)=x^{\mu}-i 2p^{\mu}\log|z|^2 +i \sum^{n=+\infty}_{n=-\infty, n\neq 0}\frac{a_n}{n}(z^{-n}+\b z^{-n}),\\
&&[x^{\mu},p^{\nu}]=i\eta^{\mu\nu},\quad [a^{\mu}_n, a^{\nu}_m]=\eta^{\mu\nu}n\delta_{n+m,0},
\end{eqnarray}
where $\eta^{\mu\nu}$ is the constant metric in the flat $D$-dimensional space of either Euclidean or Minkowski signature.
One can define the Virasoro generators 
\begin{eqnarray}
L_n=\frac{1}{2}\sum^{\infty}_{m=-\infty}:a^{\mu}_{n-m}a_{\mu m}:
\end{eqnarray}
such that $a^{\mu}_{0}=2p^{\mu}$, 
and associate with them the so-called BRST operator:\cite{brst}\cdash\cite{furuuchi}$^,$\cite{pol}
\begin{eqnarray}\label{brst}
Q=\sum^{\infty}_{n=-\infty}c_nL_{-n}+\sum^{\infty}_{m,n=-\infty}\frac{m-n}{2}:c_mc_nb_{-m-n}:-c_0,
\end{eqnarray}
where $\{c_n,b_m\}=\delta_{m,n}$ and $:\  :$ stands for Fock normal ordering. It is well known that $Q_B$ is 
nilpotent, when $D=26$. 
We note that we put the usual $\alpha'$ parameter equal to 2.\cite{pol} 
The so-called conformal vacuum $|0\rangle$, which is $sl(2,\mathbb{R})$-invariant (under the action of 
$L_0, L_{\pm 1}$), satisfies the following conditions under the action of the 
corresponding modes:

\begin{eqnarray}
&&a^{\mu}_n|0\rangle=0, \quad n\ge 0, \nonumber\\
&&b_n|0\rangle=0,\quad n\ge -1,\nonumber\\
&&c_n|0\rangle=0, \quad n > 1.
\end{eqnarray}
This leads to the relation $Q_B|0\rangle=0$. We define the ghost number operator $N_g$ by 

\begin{eqnarray}
N_g=\frac{3}{2}+\frac{c_0b_0-b_0c_0}{2}+\sum^{\infty}_{n=1}(c_{-n}b_n-b_{-n}c_n). 
\end{eqnarray} 
This constant shift (+3/2) is included to make the ghost number of $|0\rangle$ be 
equal to 0.

\noindent\underline{\bf {Bilinear operation and Lie brackets.}} In this paper, we  will meet two bilinear operations $[\cdot,\cdot]$, $[\cdot,\cdot]_h$. The first one,
without a subscript,  denotes 
the Lie bracket in the given finite-dimensional Lie algebra $\mathfrak{g}$ and the second one,
 with subscript $h$, denotes the graded antisymmetric bilinear operation 
in the homotopy Lie superalgebra. \\
\underline{\bf {Operations on differential forms.}} We will use three types of operators 
acting on differential forms with values in some finite-dimensional reductive Lie algebra 
$\mg$. 
The first one is the de Rham operator $\ud$. The second one is the {\it Maxwell operator} 
$\mathfrak{m}$, which maps 1-forms to 1-forms. Say, if $\mA=A_{\mu}dx^{\mu}$ is a 1-form, then 
\begin{eqnarray}
\m\mA=(\p_{\mu}\p^{\mu}A_{\nu}-\p_{\nu}\p_{\mu}A^{\mu})dx^{\nu}, 
\end{eqnarray}
where indices are raised and lowered w.r.t. 
the metric $\eta^{\mu\nu}$. The third operator maps 1-forms to 0-forms, this 
is the operator of divergence $\di$. For a given 1-form $\mA$, 
\begin{eqnarray}
\di\mA=\p_{\mu}A^{\mu}. 
\end{eqnarray}
For $\mg$-valued 1-forms, one can also define the following (anti)symmetric bilinear and three-linear operations:

\begin{eqnarray}
(\mA,\mB)&=&(A_{\mu},B^{\mu})_K,\nonumber\\
\{\mathbf{A},\mB\}&=&([A_{\mu},\p^{\mu}B_{\nu}]+[B_{\mu},\p^{\mu}A_{\nu}]+[\p_{\nu}A_{\mu}, B^{\mu}] \nonumber\\
&&+[\p_{\nu}B_{\mu}, A^{\mu}]+\p^{\mu}[A_{\mu},B_{\nu}]+\p^{\mu}[B_{\mu},A_{\nu}])dx^{\nu},
\end{eqnarray}
\begin{eqnarray}
\mA\cdot\mW&=&[A^{\mu},W_{\mu}],\nonumber\\
\{\mA,\mB,\mC\}&=&([A_{\mu},[B^{\mu},C_{\nu}]]+[B_{\mu},[A^{\mu},C_{\nu}]]\nonumber\\
&&+[C_{\mu},[B^{\mu},A_{\nu}]]+[B_{\mu},[C^{\mu},A_{\nu}]]\nonumber\\
&&+[A_{\mu},[C^{\mu},B_{\nu}]]+[C_{\mu},[A^{\mu},B_{\nu}]])dx^{\nu},
\end{eqnarray}
where $(\cdot,\cdot)_K$ is the canonical invariant form on the Lie algebra $\mg$.

\section{Yang-Mills Chain Complex and Homotopy Lie Algebra} 
\subsection{Chain Complex} 
Let us consider some finite-dimensional reductive Lie algebra $\mg$ and the following states of open SFT
\begin{eqnarray}\label{f0}
&&\rho_{u}=u(x)|0\rangle, \quad \phi_{\mathbf{A}}=(-ic_1A_{\mu}(x)a_{-1}^{\mu}-c_0\p_{\mu}A^{\mu}(x))|0\rangle,\nonumber\\ 
&&\psi_{\mathbf{W}}=-ic_1c_0 W_{\mu}(x)a_{-1}^{\mu}|0\rangle, \quad \chi_{a}=2c_1c_0 c_{-1} a(x)|0\rangle
\end{eqnarray} 
 associated with  $\mg$-valued functions $u(x)$, $a(x)$ and 1-forms 
$\mathbf{A}=A_{\mu}(x)dx^{\mu}, 
\mathbf{W}=W_{\mu}(x)dx^{\mu}$.  
It is easy to check that the resulting space, spanned by the states like \rf{f0}, is invariant under the action of the BRST operator,
 moreover, the following proposition holds.

\begin{proposition}
\cite{ym} {\it Let the space $\mathcal{F}_{\mg}$ be spanned by all possible states of the form \rf{f0}.  
Then we have a chain complex:
\begin{eqnarray}\label{complex}
0\to\mg\xrightarrow{i}\mathcal{F}_{\mg}^{0}\xrightarrow{Q}\mathcal{F}_{\mg}^{1}
\xrightarrow{Q}\mathcal{F}_{\mg}^{2}\xrightarrow{Q}\mathcal{F}_{\mg}^{3}\to 0,
\end{eqnarray}
where $\mathcal{F}_{\mg}^{i}$ (i=0,1,2,3) is a subspace of $\mathcal{F}$ 
corresponding to the ghost number $i$ 
and $Q$ is the BRST operator \rf{brst}.}
\end{proposition}
\begin{proof}
 Actually, it is easy to see that we have the following formulas:
\begin{eqnarray}\label{qact}
Q\rho_{u}=2\phi_{\ud u},\quad Q\phi_{\mathbf{A}}=2\psi_{\m \mA},\quad Q\psi_{\mathbf{W}}=-\chi_{\di \mW}, 
\quad Q\chi_{a}=0.
\end{eqnarray}
\break Then the statement can be easily obtained. 
\end{proof}

\begin{remark}
 From \rf{qact}, one can see that in the case where $\mg$ is Abelian,
the first cohomology module $H^1_{Q}(\mathcal{F})$ can be identified 
with the space of abelian gauge fields, satisfying the Maxwell equations modulo gauge transformations. 
\end{remark}

\noindent 
Now, we construct another realization of this chain complex which appears to be quite useful. Namely, we 
notice that acting on $\mF_{\mg}$, 
we use only the $sl(2,\mathbb{R})$ part of the operator $Q$, so one can reduce $Q$ to the usual Chevalley 
differential for the $sl(2,\mathbb{R})$ algebra generated by $L_{\pm 1}, L_0$, i.e. one can reduce $Q$ to the 
following operator:
\begin{eqnarray}\label{chev}
Q=\sum^{1}_{n=-1}c_nL_{-n}-c_0c_1b_1+c_0c_{-1}b_{-1}-2c_{-1}c_1b_0.
\end{eqnarray}
Moreover, one can easily see that $Q$ can be reduced further, to be presented as a differential 
operator
\begin{eqnarray}\label{diff}
Q=\sum^{1}_{n=-1}c_ns_{-n}-2c_{-1}c_1\frac{\p}{\p c_0},
\end{eqnarray}
where 
\begin{eqnarray}
s_0=-2\frac{\p^2}{\p x^{\mu}\p x_{\mu}},\quad s_1=-i2\frac{\p^2}{\p x^{\mu}\p q_{\mu}},\quad s_{-1}=-i2q^{\mu}\frac{\p}{\p x^{\mu}},
\end{eqnarray} 
using the usual differential operator representation for the Heisenberg algebra, 
\begin{eqnarray}
a^{\mu}_0=2p^{\mu}\to -2i\frac{\p}{\p x^{\mu}},\quad a^{\mu}_{-1}\to q^{\mu}, \quad 
a^{\mu}_{1}\to \frac{\p}{\p q^{\mu}}, \quad b_{n}\to\frac{\p}{\p c_n}
\end{eqnarray} 
and eliminating all terms from $Q$, which act as zero. Therefore, it is reasonable to write $\rho_u$, 
$\phi_{\mA}$, $\psi_{\mW}$, $\chi_a$, which span the space of our chain complex, as the following Lie 
algebra-valued functions of $c_n$, $x^{\mu}$ and $q^{\nu}$:
\begin{eqnarray}
&&\rho_{u}=u(x), \quad \phi_{\mathbf{A}}=-ic_1A_{\mu}(x)q^{\mu}-c_0\p_{\mu}A^{\mu}(x),\nonumber\\ 
&&\psi_{\mathbf{W}}=-ic_1c_0 W_{\mu}(x)q^{\mu}, \quad \chi_{a}=2c_1c_0 c_{-1} a(x).
\end{eqnarray}  
We note that the ghost number operator can be reduced to 
\begin{eqnarray}
N_g=\sum^{n=1}_{n=-1}c_n\frac{\p}{\p c_n}.
\end{eqnarray}
Our next task is to define an inner product on the chain complex $\mF_{\mg}$ which behaves in a reasonable way under 
the action of the differential. First of all, for any vector 
$\Psi\in \mF_{\mg}$, which has the explicit form
\begin{eqnarray}
\Psi=\rho_{u}+\phi_{\mathbf{A}}+\psi_{\mathbf{W}}+\chi_{a}
\end{eqnarray}  
for some $u, \mA, \mW, a$, 
we define the ``conjugate'' differential operator:\footnote{The reader with experience in CFT 
can see that it is nothing but the BPZ conjugate state.}
\begin{eqnarray}
&&\Psi^*=u(x)-ic_{-1}A_{\mu}(x)\frac{\p}{\p q_{\mu}}+c_0\p_{\mu}A^{\mu}(x)-\nonumber\\
&&ic_{-1}c_0 W_{\mu}(x)\frac{\p}{\p q^{\mu}}-2c_{-1}c_0 c_{1} a(x).
\end{eqnarray}  
Now, we are ready to define the pairing.

\begin{definition} 
Consider two elements $\Phi, \Psi$ of $\mF_{\mg}$. Their 
pairing $\langle\Psi,\Phi\rangle$ is defined by the following formula:
\begin{eqnarray}\label{ip}
\langle\Psi,\Phi\rangle=\int d^Dx\int dc_{-1}dc_0dc_{1}(\Psi^*,\Phi)_K(x,c_i),
\end{eqnarray} 
where $(\cdot,\cdot)_K$ is the canonical invariant form on $\mg$ and 
the integral over $ c_{-1},\ c_0,\ c_{1} $ is the standard Berezin integral.
\end{definition}

In the case when $\Psi=\rho_{u}+\phi_{\mathbf{A}}+\psi_{\mathbf{U}}+\chi_{a}$ and 
$\Phi=\rho_{v}+\phi_{\mathbf{B}}+\psi_{\mathbf{V}}+\chi_{b}$, the pairing is given by
\begin{eqnarray}
&&\langle\Psi,\Phi\rangle=\int d^Dx((\mA,\mathbf{V})(x)+(\mU,\mB)(x)-\nonumber\\
&&2(u(x),b(x))_K
-2(a(x),v(x))_K).
\end{eqnarray} 
Now, we formulate as a proposition how this inner product behaves under the action of the differential 
$Q$. 

\begin{proposition}
 Let $\Phi, \Psi\in \mF_{\mg}$ be of ghost numbers $n_{\Phi}$, 
$n_{\Psi}$. Then, the following relation holds:
\begin{eqnarray}
\langle Q\Phi, \Psi\rangle=-(-1)^{n_{\Phi}n_{\Psi}}\langle Q\Psi, \Phi\rangle.
\end{eqnarray}
\end{proposition}

\begin{proof}
It is easy to see that it is enough to show that the following relations hold:
\begin{eqnarray}\label{rela}
\langle Q\psi_{\mW}, \rho_u\rangle=-\langle Q \rho_u, \psi_{\mW} \rangle, \quad 
\langle Q\phi_{\mA}, \phi_{\mB}\rangle=\langle Q\phi_{\mB}, \phi_{\mA}\rangle.
\end{eqnarray} 
The proof is straightforward:
\begin{eqnarray}
\langle Q\psi_{\mW}, \rho_u\rangle&=&\int 2(\p_{\mu}W^{\mu},u)_K=-\int 2(W^{\mu},\p_{\mu}u)_K\nonumber\\
&=&-2\langle \rho_{\ud u}, \psi_{\mW}\rangle=-\langle Q \rho_u, \psi_{\mW} \rangle,\nonumber\\
\langle Q\phi_{\mA}, \phi_{\mB}\rangle&=&\int 2(\p_{\mu}\p^{\mu}A_{\nu}, B^{\nu})_K\\
&=&\int 2(\p_{\mu}\p^{\mu}B_{\nu}, A^{\nu})_K=\langle Q\phi_{\mB}, \phi_{\mA}\rangle. \nonumber\break
\end{eqnarray} 
\end{proof}

\subsection{Homotopy Lie Algebra and Yang-Mills action} 
Now, we construct the graded bilinear and 
three-linear operations on the space of our chain complex. 

\begin{definition}\cite{ym}  We define the bilinear operation  
\begin{eqnarray}
[\cdot, \cdot]_h: \mathcal{F}^i_{\mathfrak{g}}\otimes \mathcal{F}^j_{\mathfrak{g}}\to \mathcal{F}^{i+j}_{\mathfrak{g}},
\end{eqnarray}
which is graded (w.r.t. to the ghost number) as an antisymmetric bilinear operation 
by the following relations on the elements of 
$\mathcal{F}_{\mathfrak{g}}$: 
\begin{eqnarray}
&&[\rho_{u},\rho_{v}]_h=2\rho_{[u,v]}, \quad
[\rho_{u},\phi_{\mathbf{A}}]_h=2\phi_{[u,\mathbf{A}]}, \quad 
[\rho_{u},\psi_{\mathbf{W}}]_h=2\phi_{[u,\mathbf{W}]}, \nonumber\\
&&[\rho_u, \chi_a]_h=2\chi_{[u,a]},\quad[\phi_{\mathbf{A}},\phi_{\mathbf{B}}]_h=2\phi_{\{\mathbf{A},\mB\}}, \quad [\phi_{\mathbf{A}},\psi_{\mathbf{W}}]_h= 
-\chi_{\mathbf{A}\cdot\mW},
\end{eqnarray}
where $\rho_u$, $\rho_v \in \mF^0_{\mathfrak{g}}$, $\phi_{\mathbf{A}}$, $\phi_{\mathbf{B}}\in \mF^1_{\mathfrak{g}}$, 
$\psi_{\mathbf{W}}\in \mF^2_{\mathfrak{g}}$, $\chi_a\in \mF^3_{\mathfrak{g}}$.
\end{definition}
\begin{definition}\cite{ym}  The operation  
\begin{eqnarray}
\label{3lin}[\cdot, \cdot, \cdot]_h: \mathcal{F}^i_{\mathfrak{g}}\otimes \mathcal{F}^j_{\mathfrak{g}}\otimes \mathcal{F}^k_{\mathfrak{g}}\to 
\mathcal{F}^{i+j+k-1}_{\mathfrak{g}}
\end{eqnarray}
is defined to be nonzero only when all arguments lie in $\mF_{\mg}^1$. For $\phi_{\mathbf{A}}$, 
$\phi_{\mathbf{B}}$, $\phi_{\mathbf{C}}$$\in \mF_{\mg}^1$, we have
\begin{eqnarray}
[\phi_{\mathbf{A}},\phi_{\mathbf{B}} ,\phi_{\mathbf{C}} ]_h=2\psi_{\{\mA,\mB,\mC\}}.
\end{eqnarray}
\end{definition}
\begin{remark}
Here, we note that the bilinear operation, defined in this subsection, corresponds to the lowest orders in $\alpha'$ 
of that introduced in Ref.~ \refcite{zeit3} (we will return to this in Subsec. 2.3). 
\end{remark}
We claim that these graded antisymmetric multilinear 
operations satisfy the relations of a homotopy Lie algebra. Namely, the following proposition holds.

\begin{proposition}\cite{ym} {\it Let $a_1,a_2, a_3, b, c$ $\in$ $\mF_{\mathfrak{g}}$ be of ghost numbers 
$n_{a_1}$, $n_{a_2}$, $n_{a_3}$, $n_b$, $n_c$ correspondingly. Then the following relations hold:}
\begin{eqnarray}\label{rel}
&&Q[a_1,a_2]_h=[Q a_1,a_2]_h+(-1)^{n_{a_1}}[a_1,Q a_2]_h,\nonumber\\
&&Q[a_1,a_2, a_3]_h+[Q a_1,a_2, a_3]_h+(-1)^{n_{a_1}}[a_1,Q a_2, a_3]_h+\nonumber\\
&&(-1)^{n_{a_1}+n_{a_2}}[ a_1, a_2, Q a_3]_h+[a_1,[a_2, a_3]_h]_h-[[a_1,a_2]_h, a_3]_h-\nonumber\\
&&(-1)^{n_{a_1}n_{a_2}}[a_2,[a_1, a_3]_h]_h=0,\nonumber\\
&&[b,[a_1,a_2, a_3]_h]_h-(-1)^{n_b(n_{a_1}+n_{a_2}+n_{a_3})}[a_1,[a_2, a_3, b]_h]_h+\nonumber\\
&&(-1)^{n_{a_2}(n_{b}+n_{a_1})}[a_2,[b,a_1, a_3]_h]_h-(-1)^{n_{a_3}(n_{a_1}+n_{a_2}+n_{b})}
[a_3,[b, a_1,a_2]_h]_h\nonumber\\
&&=[[b,a_1]_h,a_2, a_3]_h+(-1)^{n_{a_1}n_{b}}[a_1,[b,a_2]_h, a_3]_h+\nonumber\\
&&(-1)^{(n_{a_1}+n_{a_2})n_{b}}[a_1,a_2, [b,a_3]_h]_h,\nonumber\\
&&[[a_1,a_2, a_3]_h,b,c]_h=0.
\end{eqnarray}
\end{proposition} 
The proof is given in the Appendix.

\begin{remark} Denoting $d_0=Q$, $d_1=[\cdot, \cdot]_h$, $d_2=[\cdot, \cdot, \cdot]_h$, the relations \rf{rel}, together 
with the condition $Q^2=0$, can be summarized 
in the following way: 
\begin{eqnarray}\label{d2}
D^2=0,
\end{eqnarray}
where $D=d_0+\theta d_1+\theta^2d_2$ . Here, $\theta$ is some formal parameter anticommuting with $d_0$ and $d_2$. 
We recall that $d_0$ raises the ghost number, by $1$, $d_1$ leaves it unchanged and $d_2$ lowers it by 1. 
Therefore, $d_0,d_2$ are odd elements as well as the parameter $\theta$, but $d_1$ is even. Hence, \rf{d2} gives the 
following relations
\begin{eqnarray}
&&d_0^2=0,\quad d_0d_1-d_1d_0=0, \quad d_1d_1+d_0d_2+d_2d_0=0,\nonumber\\ 
&&d_1d_2-d_2d_1=0, \quad d_2d_2=0,
\end{eqnarray}
which are in agreement with \rf{rel}.
\end{remark}
Now, we will define the following multilinear forms on the complex $\mF_{\mg}$ which appear to be very 
useful for the construction of the Yang-Mills action.
\begin{definition}  For any $a_1, a_2, a_3, a_4\in \mF_{\mg}$, one can define the  n-linear 
forms (n=2,3,4)
\begin{equation}
\{\cdot,..., \cdot\}_h: \mF_{\mg}\otimes ...\otimes \mF_{\mg}\to \mathbb{C} 
\end{equation}
in the following way:
\begin{eqnarray}
\{a_1,a_2\}_h&=&\langle Qa_1, a_2\rangle, \nonumber\\
 \{a_1,a_2,a_3\}_h&=&\langle [a_1, a_2]_h,a_3\rangle,\nonumber\\
\{a_1,a_2,a_3, a_4\}_h&=&\langle [a_1, a_2,a_3]_h,a_4\rangle.
\end{eqnarray}
\end{definition}
These bilinear operations satisfy the following remarkable property.

\begin{proposition} The multilinear forms, introduced in Definition 2.5, are graded as antisymmetric, i.e.
\begin{eqnarray}
\{a_1,...,a_i,a_{i+1},..., a_n\}_h=-(-1)^{n_{a_i}n_{a_{i+1}}}\{a_1,...,a_{i+1},a_{i},..., a_n\}_h,
\end{eqnarray}
where $n_{a_i}$ denotes the ghost number of $a_i$.
\end{proposition}
The proof is given in the Appendix.

Now, we are ready to formulate the Yang-Mills action as a homotopy Chern-Simons theory.

\begin{proposition} The Yang-Mills action 
\begin{eqnarray}
S_{YM}=1/2\int d^D x(F_{\mu\nu}(x),F^{\mu\nu}(x))_K, \quad F_{\mu\nu}=\p_{\mu}A_{\nu}-\p_{\nu}A_{\mu}+[A_{\mu},A_{\nu}],
\end{eqnarray}
can be written as follows:
\begin{eqnarray}\label{cs}
&&S_{YM}=-\sum^4_{n=2}\frac{1}{n!}\{\phi_{\mA}^n\}_h=\nonumber\\
&&-\frac{1}{2}\langle Q\phi_{\mA},\phi_{\mA}\rangle-\frac{1}{6}\{\phi_{\mA},\phi_{\mA},\phi_{\mA}\}_h\nonumber\\
&&-\frac{1}{24}
\{\phi_{\mA},\phi_{\mA},\phi_{\mA}, \phi_{\mA}\}_h.
\end{eqnarray}
\end{proposition}
\begin{proof} 
Actually, from the definition of the brackets, one can see that
\begin{eqnarray}
&&Q\phi_{\mA}=2\psi_{\mW_1}, \quad W_{1\mu}=\p_{\nu}\p^{\nu}A_{\mu}-\p_{\mu}\p^{\nu}A_{\nu},\nonumber\\
&&[\phi_{\mA},\phi_{\mA}]_h=4\psi_{\mW_2}, \quad W_{3\mu}=[\p_{\nu}A^{\nu},A_{\mu}]+
2[A^{\nu},\p_{\nu}A_{\mu}]-[A^{\nu},\p_{\mu}A_{\nu}],\nonumber\\
&&\label{gmc}[\phi_{\mA},\phi_{\mA},\phi_{\mA}]_h=12\psi_{\mW_3},\quad W_{3\mu}=[A_{\nu}, [A^{\nu},A_{\mu}]].
\end{eqnarray}
Therefore, 
\begin{eqnarray}
&&\frac{1}{2}\langle Q\phi_{\mA},\phi_{\mA}\rangle=\langle \int d^D x(A^{\nu}(x),\p_{\mu}\p^{\mu}A_{\nu}(x)-\p_{\nu}
\p^{\mu}A_{\mu}(x))_K\rangle=\nonumber\\
&&-1/2\int d^D x(\p_{\mu}A_{\nu}(x)-\p_{\nu}A_{\mu}(x), \p_{\mu}A_{\nu}(x)-\p_{\nu}A_{\mu}(x))_K,\nonumber\\
&&\frac{1}{6}\{\phi_{\mA},\phi_{\mA},\phi_{\mA}\}_h=
-\int d^D x(\p^{\nu}A^{\mu}(x)-\p^{\mu}A^{\nu}(x))[A_{\nu}(x), A_{\mu}(x)],\nonumber\\
&&\frac{1}{24}\{\phi_{\mA},\phi_{\mA},\phi_{\mA}, \phi_{\mA}\}_h=-\frac{1}{2}\int d^D x ([A_{\mu}(x),A_{\nu}(x)], 
[A^{\mu}(x),A^{\nu}(x)])_K.
\end{eqnarray}
Thus, we see that 
\begin{eqnarray}
&&\frac{1}{2}\langle Q\phi_{\mA},\phi_{\mA}\rangle+\frac{1}{6}\{\phi_{\mA},\phi_{\mA},\phi_{\mA}\}_h+\nonumber\\
&&\frac{1}{24}
\{\phi_{\mA},\phi_{\mA},\phi_{\mA}, \phi_{\mA}\}_h=-\frac{1}{2}\int d^D x(F_{\mu\nu}(x),F^{\mu\nu}(x))_K.
\end{eqnarray}
\break This finishes the proof. 
\end{proof}
Varying the action \rf{cs} with respect to $\phi_{\mA}$, one sees that the resulting equations of motion are the 
generalized Maurer-Cartan equations, which obviously coincide with Yang-Mills equations. Moreover, 
it appears that the gauge transformations 
coincide with the transformations which preserve the Maurer-Cartan equations. This can be summarized in the following proposition.

\begin{proposition}\cite{ym} 
Let $\phi_{\mA}$ be the element of $\mF^1_{\mathfrak{g}}$ associated with 1-form $\mA=A_{\mu}dx^{\mu}$ and 
$\rho_u$ be the element of $\mF^0_{\mathfrak{g}}$ associated with Lie algebra-valued function $u(x)$. 
Then, the Yang-Mills equations for $\mA$ and its infinitesimal gauge transformations
\begin{eqnarray}\label{ym}
\p_{\mu}F^{\mu\nu}+[A_{\mu},F^{\mu\nu}]=0, \quad \label{gt}A_{\mu}\to A_{\mu}+\epsilon(\p_{\mu}u+[A_{\mu},u])
\end{eqnarray}
can be rewritten as follows:
\begin{eqnarray}\label{mc}
&&Q\phi_{\mA}+\frac{1}{2!}[\phi_{\mA},\phi_{\mA}]_h+\frac{1}{3!}[\phi_{\mA},\phi_{\mA},\phi_{\mA}]_h=0,\\
&&\phi_{\mA}\to \phi_{\mA}+\frac{\epsilon}{2}(Q\rho_u +[\phi_{\mA},\rho_u]_h).
\end{eqnarray}

\end{proposition}

\subsection{OPE origin of the bilinear operation $[\cdot, \cdot]_h$}

 In this subsection, we  show 
how we found the bilinear operation $[\cdot, \cdot]_h$. 

We consider the open string theory on the disk, conformally mapped to the upper half-plane. 
The operator products between the coordinate fields are the following (we include 
the usual $\alpha'$-coefficient in the operator product):

\begin{eqnarray}\label{ope}
X^{\mu}(z_1)X^{\nu}(z_2)\sim -\frac{\alpha'}{2}\eta^{\mu\nu}\log|z_1-z_2|^2-\frac{\alpha'}{2}\eta^{\mu\nu}\log|z_1-\b z_2|^2.
\end{eqnarray}
The states \rf{f0} correspond to the following operators: 
 
\begin{eqnarray}\label{op}
u(X), \quad cA_{\mu}(X)\p X^{\mu}-\frac{\alpha'}{2}\p c\p_{\mu}A^{\mu}(X),\quad c\p c W_{\mu}(X)\p X^{\mu}, \quad c\p c \p^2 c a(X).
\end{eqnarray}
Let $A(z)$, $B(z)$ be any two operators. Let us consider the expression 
\begin{eqnarray}
[A(t+\epsilon),B(t)],
\end{eqnarray}
where $t$ lies on the real axis and $[\cdot,\cdot]$ means the commutator in Lie algebra $\mg$. 
Due to \rf{ope}, this object is  the series in $\epsilon$ and $\log(\epsilon/\mu)$, and therefore this allows us to define 
the following operation.

\begin{definition}\cite{zeit3} {\it For any two operators $A(z), B(z)$ we define 
a bilinear operation:
\begin{eqnarray}\label{r}
R(A,B)(t)=\mathcal{P}[A(t+\epsilon),B(t)]-(-1)^{n_A n_{B}}
\mathcal{P}[B(t+\epsilon),A(t)],
\end{eqnarray}
where $\mathcal{P}$ is the projection on the $\epsilon^0(\log(\epsilon/\mu))^0$ term, $t$ lies on the real axis and $n_A, n_B$ are the ghost numbers of 
$A,B$ correspondingly.}
\end{definition}
By means of straightforward calculation,  substituting the operators \rf{op} in \rf{r}, one finds that 
the operation $R$ at the lowest orders in $\alpha'$ reproduces the 
bilinear operation $[\cdot, \cdot]_h$. Moreover, due to the following proposition,
 one obtains the first nontrivial relation of the homotopy algebra, namely 
the commutation relation between the operator $Q$ and $[\cdot, \cdot]_h$.

\begin{proposition}\cite{zeit3}  Let $A(t),B(t)$ be some operators of ghost numbers $n_A, n_B$. Then 

\begin{eqnarray}
[Q,R(A,B)]=R([Q,A],B)+(-1)^{n_{A}}R(A,[Q,B]),
\end{eqnarray}

\noindent where Q is BRST operator.

\end{proposition}

\noindent For further information on the subject see Ref.~ \refcite{zeit3}.

\section{Fields, Antifields and BV Yang-Mills}

\subsection{Fermionic degrees of freedom and total ghost number} 
In order to introduce ghosts, antifields, i.e. the fermion degrees of freedom, we 
consider the tensor product of our chain complex $(\mF_{\mg},Q)$ with some Grassmann algebra $A$. 
We assume that $A$ is $\mathbb{Z}$-graded: 
$A=\oplus_{i\in \mathbb{Z}}A^i$, and if $\lambda^i\in A^i$ and $\lambda^j\in A^j$, 
then  $\lambda^i\lambda^j=(-1)^{ij}\lambda^j\lambda^i$.
 
Moreover, we introduce the following notation: if $\lambda\in A^i$, we will say that $\lambda$ is of 
{\it target space ghost number} $i$.
Therefore, it is reasonable to introduce the gradation w.r.t. the {\it total ghost number}, which 
is equal to the sum of the {\it worldsheet ghost number}, generated by the operator $N_g$, 
and this target space ghost number on the space $\mathcal{H}_{\mg}=\mF_{\mg}\otimes A$. Hence,
 if the element $\Phi\in \mathcal{H}^n_{\mg}$ (i.e. of total ghost number $n$), 
which is written in the form $\sum_s\Phi_s\otimes\xi_s$ such that $\Phi_s\in \mF_{\mg}$ of ghost number $n^w_s$ and $\xi_s\in A$ 
of ghost number $n^t_s$, 
then $n^w_s+n^t_s=n$ for all $s$. In the following, to simplify the notation, 
we will refer to the total ghost number simply as the ghost number.  
In such a way, one can consider a new chain complex, which is now infinite:

\begin{eqnarray}\label{infcomplex}
...\xrightarrow{Q}\mathcal{H}_{\mg}^{-1}\xrightarrow{Q}\mathcal{H}_{\mg}^{0}\xrightarrow{Q}\mathcal{H}_{\mg}^{1}
\xrightarrow{Q}\mathcal{H}_{\mg}^{2}\xrightarrow{Q}...,
\end{eqnarray}

\noindent where $\mQ=Q\otimes 1$. The space $\mathcal{H}_{\mg}$ is therefore spanned by the elements of the form $\rho_u$, $\phi_{\mA}$, $\psi_{\mW}$, 
$\chi_a$, where they are associated with the functions and 1-forms, which take values in $\mg \otimes A$. 

\vspace{10mm}
\subsection{Algebraic operations and multilinear forms on the complex $\mathcal{H}_{\mg}$}
 Now, we can construct the algebraic structures, defined for the complex $\mF_{\mg}$ in Sec. 2, in the 
case of complex $\mathcal{H}_{\mg}$. 
\begin{definition} 
 Let $\Phi_i\in \mathcal{H}_{\mg}$ (i=1,2,3) such that 
$\Phi_i$=$\sum_s \Phi_i^s\otimes \xi_i^s$,  where  $\Phi_i^s\in \mF_{\mg}$ and $\xi_i^s\in A$. Then, one can define the 
multilinear algebraic operations:

\begin{eqnarray}
&&\langle \cdot, \cdot\rangle: \mathcal{H}_{\mg}\otimes \mathcal{H}_{\mg}\to A,\nonumber\\
&&[\cdot, \cdot]_h: \mathcal{H}_{\mg}\otimes \mathcal{H}_{\mg}\to \mathcal{H}_{\mg},\nonumber\\
&&[\cdot, \cdot, \cdot]_h: \mathcal{H}_{\mg}\otimes \mathcal{H}_{\mg}\otimes \mathcal{H}_{\mg}\to \mathcal{H}_{\mg}
\end{eqnarray}
by means of the  expressions:

\begin{eqnarray}
&&\langle \Phi_1, \Phi_2\rangle=\sum_{s,s'}\langle \Phi_1^s, \Phi_2^{s'}\rangle \otimes\xi_1^s\xi_2^{s'}(-1)^{n_{\xi_1^s}n_{\Phi_2^{s'}}},\nonumber\\
&&[\Phi_1, \Phi_2]_h=\sum_{s,s'}[\Phi_1^s, \Phi_2^{s'}]_h\otimes\xi_1^s\xi_2^{s'}(-1)^{n_{\xi_1^s}n_{\Phi_2^{s'}}},\nonumber\\
&&[\Phi_1, \Phi_2, \Phi_3]_h=\nonumber\\
&&\sum_{s,s',s''}[\Phi_1^s, \Phi_2^{s'}, \Phi_3^{s''}]_h\otimes\xi_1^s\xi_2^{s'}\xi_3^{s''}(-1)^{n_{\xi_1^s}(n_{\Phi_2^{s'}}+
n_{\Phi_3^{s''}})}(-1)^{n_{\xi_2^s}n_{\Phi_3^{s''}}},
\end{eqnarray}
where $n$ denotes the ghost number.
\end{definition}
It is easy to see that the operation $\langle \cdot, \cdot \rangle$  is graded as symmetric on the space $\mH_{\mg}$ 
w.r.t. the ghost number, i.e. for $\Phi, \Psi\in \mH_{\mg}$ 
\begin{eqnarray}
\langle \Phi, \Psi \rangle=(-1)^{n_{\Phi}n_{\Psi}}\langle \Psi, \Phi \rangle.
\end{eqnarray}
Similarly, one can show that on the space $\mH_{\mg}$, $[\cdot, \cdot]_h$ and $[\cdot, \cdot, \cdot]_h$  are graded as 
antisymmetric w.r.t. the ghost number and satisfy, together with $\mQ$, the relations of the homotopy Lie algebra 
\rf{rel}. 
Now, we give a definition which is analogous to Definition 2.4.

\begin{definition}
 For any $a_1, a_2, a_3, a_4\in \mH_{\mg}$ one can define the  n-linear 
forms (n=2,3,4)
\begin{equation}
\{\cdot,..., \cdot\}_h: \mH_{\mg}\otimes ...\otimes \mH_{\mg}\to A 
\end{equation}
in the following way:
\begin{eqnarray}
&&\{a_1,a_2\}_h=\langle \mQ a_1, a_2\rangle, \quad \{a_1,a_2,a_3\}_h=\langle [a_1, a_2]_h,a_3\rangle,\nonumber\\
&&\{a_1,a_2,a_3, a_4\}_h=\langle [a_1, a_2,a_3]_h,a_4\rangle.
\end{eqnarray}
\end{definition}
Using Proposition 2.4 and Definition 3.1, we find that the operations $\{\cdot,..., \cdot\}_h$ are graded antisymmetric on 
$\mH_{\mg}$. 

Finally, we mention that the properties of antisymmetric multilinear operations 
are related to the action of the element of Grassmann algebra $A$ on them. 
First of all, we have the natural right action of an element $a\in A$ on 
$\Phi\in \mH_{\mg}$, namely $\Phi\cdot a=\Phi(1\otimes a)$. Due to Definitions 3.1 and 3.2, we have
\begin{eqnarray}
&&[\Phi_1,...,\Phi_i\cdot a+ \Phi'_{i}\cdot a',...\Phi_k]_h=[\Phi_1,...,\Phi_i,...\Phi_k]_h\cdot a(-1)^{n_a(n_{\Phi_{i+1}}+...+n_{\Phi_k})}\nonumber\\
&&+[\Phi_1,...,\Phi'_i,...\Phi_k]_h\cdot a'(-1)^{n_{a'}(n_{\Phi_{i+1}}+...+n_{\Phi_k})},\nonumber\\
&&\{\Phi_1,...,\Phi_i\cdot a+ \Phi'_{i}\cdot a',...\Phi_m\}_h=\nonumber\\
&&\{\Phi_1,...,\Phi_i,...\Phi_m\}_h\cdot a(-1)^{n_a(n_{\Phi_{i+1}}+...+n_{\Phi_m})}\nonumber\\
&&+\{\Phi_1,...,\Phi'_i,...\Phi_m\}_h\cdot a'(-1)^{n_{a'}(n_{\Phi_{i+1}}+...+n_{\Phi_m})},
\end{eqnarray}
where $k=2,3$, $m=2,3,4$ and $\Phi_r\in \mH_{\mg}$ for all $r$. 

\subsection{Ghosts, antifields and BV Yang-Mills} 
One of the main characters throughout this subsection is the element $\Phi\in\mH_{\mg}$ of ghost number 1. 
Using the realization of $\mF_{\mg}$, which we have constructed in Subsec. 2.1, one can write 
the expression for such an element as follows:
\begin{eqnarray}\label{field}
&&\Phi(\omega, \omega^*, \mA, \mA^*)=\rho_{\omega}+\phi_{\mathbf{A}}-\psi_{\mathbf{A^*}}-1/2\chi_{\omega^*}=\nonumber\\
&&\omega(x)-ic_1A_{\mu}(x)q^{\mu}-c_0\p_{\mu}A^{\mu}(x)+
ic_1c_0 A^*_{\mu}(x)q^{\mu}-c_1c_0 c_{-1} \omega^*(x),
\end{eqnarray}
where the target space ghost numbers of $\omega, \omega^*, \mA, \mA^*$ are $1$, $-2$, $0$, $-1$ correspondingly. 
In this case, the analogue of Proposition 2.5. looks as follows.

\begin{proposition}
Consider $\Phi=\Phi(\omega, \omega^*, \mA, \mA^*)\in \mH^1_{\mg}$ with the notation as 
in \rf{field}. Then, the Homotopy Chern-Simons action
\begin{eqnarray}
&&S_{HCS}=-\sum^4_{n=2}\frac{1}{n!}\{\Phi^n\}_h=\nonumber\\
&&-\frac{1}{2}\langle \mQ\Phi,\Phi\rangle-\frac{1}{6}\{\Phi,\Phi,\Phi\}_h-\frac{1}{24}
\{\Phi,\Phi,\Phi, \Phi\}_h
\end{eqnarray}
coincides with the BV Yang-Mills action
\begin{eqnarray}
&&S_{BVYM}=\nonumber\\
&&S_{YM}[\mA]+2\int d^D x(D_{\mu}\omega(x),A^{*\mu}(x))_K-([\omega(x),\omega(x)],\omega^*(x))_K),
\end{eqnarray}
where, as usual, $D_{\mu}\omega=\p_{\mu}\omega+[A_{\mu},\omega]$.
\end{proposition}
\begin{proof} 
Actually, from the definition of the brackets, one can see that:
\begin{eqnarray}
&&\mQ\Phi=\mQ\phi_{\mA}+2\phi_{\ud \omega}-2\chi_{\di \mA^*}, \nonumber\\
&&[\Phi,\Phi]_h=[\phi_{\mA},\phi_{\mA}]_h+2\rho_{[\omega,\omega]}-4\phi_{[\omega, \mA]}-4\psi_{[\omega,\mA^*]}+
2\chi_{[\omega, \omega^*]} +2\chi_{\mA\cdot \mA^*},\nonumber\\
&&[\Phi,\Phi,\Phi]_h=[\phi_{\mA},\phi_{\mA},\phi_{\mA}]_h.
\end{eqnarray}
Therefore, 
\begin{eqnarray}
&&\frac{1}{2}\langle \mQ\Phi,\Phi\rangle=\frac{1}{2}\langle Q\phi_{\mA},\phi_{\mA}\rangle+\int d^Dx((A^*_{\mu},\p^{\mu}\omega)_K+
\nonumber\\
&&\int d^Dx(\p^{\mu}\omega,A^*_{\mu})_K=\frac{1}{2}\langle Q\phi_{\mA},\phi_{\mA}\rangle-2\int d^Dx(\p^{\mu}\omega(x),A^*_{\mu}(x))_K\nonumber\\
&&\frac{1}{6}\{\Phi,\Phi,\Phi\}_h=\frac{1}{6}\{\phi_{\mA},\phi_{\mA},\phi_{\mA}\}-\int d^Dx(4(A_{\mu}(x),[\omega(x),A^*_{\mu}(x)])_K-\nonumber\\
&&4(\omega,[A^{\mu}(x),A_{\mu}(x)])_K+4(A^*_{\mu}(x),[\omega(x),A_{\mu}(x)])_K\nonumber\\
&&-2(\omega^*,[\omega,\omega])_K-4(\omega,[\omega,\omega^*])_K)\nonumber\\
&&=\frac{1}{6}\{\phi_{\mA},\phi_{\mA},\phi_{\mA}\}-12\int d^Dx([A_{\mu}(x),\omega(x)],A^{*\mu}(x))_K
\nonumber\\
&&\frac{1}{24}\{\Phi,\Phi,\Phi,\Phi\}_h=\frac{1}{24}\{\phi_{\mA},\phi_{\mA},\phi_{\mA}, \phi_{\mA}\}.
\end{eqnarray}
\break Summing all these terms, we see that $S_{HCS}=S_{BVYM}$. This finishes the proof.
\end{proof}

\section{BV Formalism, Gauge Conditions and Qu\-an\-tum Theory.}
  First of all, we  shortly review the basic elements of the BV formalism. Consider the set of 
all fields $\psi_n(x)$ in some classical field theory in $D$ dimensions. Let us assign to each field $\psi_n(x)$ the $antifield$ 
$\psi^*_n(x)$, which is of opposite statistics to $\psi_n(x)$. Then, one can define the odd Laplace operator, sometimes called 
the BV Laplacian, and the BV bracket:

\begin{eqnarray}
&&\Delta_{BV}=\int d^Dx\frac{\delta}{\delta\psi_n(x)}\frac{\delta}{\delta\psi^*_n(x)},\nonumber\\
&&(F,G)_{BV}=\int d^D x(\frac{\delta_R F}{\delta\psi_n(x)}\frac{\delta_L G}{\delta\psi^*_n(x)}-
\frac{\delta_R F}{\delta\psi^*_n(x)}\frac{\delta_L G}{\delta\psi_n(x)}),
\end{eqnarray}
where $F,G$ are some functionals on the space of fields and antifields which can be either even or odd.
The $classical$ $BV $ $action$ $S^{BV}_{cl}$ is a bosonic functional, which satisfies the {\it classical BV Master equation}:

\begin{eqnarray}\label{bvcl}
(S^{BV}_{cl},S^{BV}_{cl})_{BV}=0.
\end{eqnarray}
For the gauge theories, in particular for Yang Mills with the gauge-invariant action $S[\mA]$, where $\mA$ is a gauge field, it is easy 
to construct an action satisfying Eq. \rf{bvcl}:
\begin{eqnarray}\label{gauge}
&&S_{cl}^{BV}[\mA, \omega,\mA^*, \omega^*]=\nonumber\\
&&S[\mA]+2\int d^D x(D_{\mu}\omega(x),A^{*\mu}(x))_K-([\omega(x),\omega(x)],\omega^*(x))_K).
\end{eqnarray}
Here, $\omega$ is a so-called ghost field, $\omega$ is its antifield, and $\mA^*$ is the antifield for $\mA$. 
As we have seen in the previous section for Yang-Mills gauge theory, this action can be written as a homotopy Chern-Simons theory. 

The action \rf{gauge} satisfies the so-called {\it quantum BV Master equation}:
\begin{eqnarray}
\Delta_{BV}(e^{-\frac{1}{h}S[\psi_n,\psi^*_n] })=0.
\end{eqnarray}
The quantum BV theory is defined by means of a path integral over some Lagrangian submanifold $L$ which is defined with respect to 
the odd symplectic form $\sum_n\delta\psi_n\wedge \delta\psi^*_n$, i.e. 
$\sum_n\delta\psi_n\wedge \delta\psi^*_n\Bigl\lvert_L=0$. One of the great advantages of 
this formalism is that the corresponding effective action, which we find after integrating over the appropriately chosen parts of fields and 
antifields, will also satisfy the quantum BV Master equation.  

One of the most popular Lagrangian submanifolds for the action \rf{gauge} is
\begin{eqnarray}
\omega^*=0, \quad F(\mA)=0, \quad  \mA^*=-\frac{\delta F^{\b\omega}}{\delta \mA}, 
\end{eqnarray}
where $F$ is some Lie algebra-valued constraint on gauge fields, $\b \omega$ is what is left of $\mA^*$ degree of freedom and is 
usually called the $antighost$ field,\cite{weinberg} and $F^{\b \omega}=\int d^Dx(\b \omega(x), F(x))_K$.
 
Actually, substituting the values for antifields in the action, we get 
\begin{eqnarray}\label{gf}
S^{gf}_{YM}=S_{YM}-2\int d^Dx(D_{\mu}\omega(x),\frac{\delta F^{\b\omega}}{\delta A_{\mu}(x)})_K\Bigl\lvert_{F=0},
\end{eqnarray}
which coincides with the usual gauge fixed action for quantum Yang-Mills theory.\cite{weinberg} For example, in the case of 
the Lorentz gauge, when $F=\p_{\mu}A^{\mu}$, the action reads
\begin{eqnarray}
S^{gfL}_{YM}=S_{YM}-2\int d^Dx(\b \omega(x), \p^{\mu} D_{\mu}\omega(x))_K\Bigl\lvert_{\p_{\mu}A^{\mu}=0}.
\end{eqnarray}
Remember that in our formalism, the condition $b_0\Phi(\mA, \omega,\mA^*, \omega^*)=0$, which in open SFT is known as the
{\it Siegel gauge} \cite{siegel}, is equivalent to the following conditions on fields and antifields: 
$\p_{\mu}A^{\mu}=0$, $\omega^*=0$, $\mA^{*}=0$. This is not precisely what we are willing to obtain. However, one can 
consider the object $R_{\b\omega}=1/2\langle \chi_{\b\omega},b_0\Phi \rangle$. Then, one can define  
\begin{eqnarray}
S^{gfL}_{HCS}(\Phi)=S_{HCS}(\Phi)+(S_{HCS}(\Phi),R_{\b\omega}(\Phi))_{BV}\Bigl\lvert_{b_0\Phi=0},
\end{eqnarray}
which coincides with $S^{gfL}_{YM}$.

\section{Final Remarks}
In this paper, we have reformulated BV Yang-Mills theory into homotopy Chern-Simons theory. From the very beginning, we were 
motivated by the constructions of open SFT. In the case of closed SFT, one can get the action for linearized gravity in the 
Abelian Chern-Simons-like form. 

In Refs.~ \refcite{lmz}--\refcite{zeit2} and \refcite{zeit3}, motivated by the structures from Zwiebach's closed SFT, we constructed 
the bilinear operations on 
the corresponding operators, which should be associated with some homotopy Lie algebra, and obtained the Einstein equations up to the 
second order from the associated formal Maurer-Cartan equations. 

This gives us the possibility of thinking that one can reformulate Einstein's theory of gravity in the homotopy Chern-Simons form;
 however, the number of multilinear operations in the associated homotopy algebra and, therefore in the action, can be infinite.

\section*{Acknowledgements} 
I would like to thank D. Borisov, I. B. Frenkel, M. M. Kapranov, M. Movshev, T. Pantev, M. Rocek, 
H. Sati, J. Stasheff, D. Sullivan, M. A. Vasiliev and G. Zuckerman for numerous discussions on the subject 
and also I. B. Frenkel and N. Yu. Reshetikhin for their contant encouragement and support. 

\appendix
\section{}

\noindent{\bf Proposition A.1.}\cite{ym} {\it Let $a_1,a_2, a_3, b, c$ $\in$ $\mF$ be of ghost numbers 
$n_{a_1}$, $n_{a_2}$, $n_{a_3}$, $n_b$, $n_c$ correspondingly. Then the following relations hold:
\begin{eqnarray}
&&Q[a_1,a_2]_h=[Q a_1,a_2]_h+(-1)^{n_{a_1}}[a_1,Q a_2]_h,\nonumber\\
&&Q[a_1,a_2, a_3]_h+[Q a_1,a_2, a_3]_h+(-1)^{n_{a_1}}[a_1,Q a_2, a_3]_h+\nonumber\\
&&(-1)^{n_{a_1}+n_{a_2}}[ a_1, a_2, Q a_3]_h+[a_1,[a_2, a_3]_h]_h-[[a_1,a_2]_h, a_3]_h-\nonumber\\
&&(-1)^{n_{a_1}n_{a_2}}[a_2,[a_1, a_3]_h]_h=0,\nonumber\\
&&[b,[a_1,a_2, a_3]_h]_h-(-1)^{n_b(n_{a_1}+n_{a_2}+n_{a_3})}[a_1,[a_2, a_3, b]_h]_h+\nonumber\\
&&(-1)^{n_{a_2}(n_{b}+n_{a_1})}[a_2,[b,a_1, a_3]_h]_h-(-1)^{n_{a_3}(n_{a_1}+n_{a_2}+n_{b})}
[a_3,[b, a_1,a_2]_h]_h\nonumber\\
&&=[[b,a_1]_h,a_2, a_3]_h+(-1)^{n_{a_1}n_{b}}[a_1,[b,a_2]_h, a_3]_h+\nonumber\\
&&(-1)^{(n_{a_1}+n_{a_2})n_{b}}[a_1,a_2, [b,a_3]_h]_h,
\nonumber\\
&&[[a_1,a_2, a_3]_h,b,c]_h=0.
\end{eqnarray} }
\begin{proof}
Let us start from the first relation:
\begin{eqnarray}\label{der}
Q[a_1,a_2]_h=[Q a_1,a_2]_h+(-1)^{n_{a_1}}[a_1,Q a_2]_h.
\end{eqnarray}
We begin from the case where $a_1=\rho_{u}\in \mF^0_{\mathfrak{g}}$. Then, for $a_2=\rho_v\in \mF^0_{\mathfrak{g}}$, we have

\begin{eqnarray}
Q[\rho_u,\rho_v]_h=4\phi_{\ud [u,v]}=[\rho_u,2\phi_{\ud v}]_h+[2\phi_{\ud u}, \rho_v]_h=
[Q\rho_u,\rho_v]_h+[\rho_u,Q\rho_v]_h.
\end{eqnarray}
Let  $a_2=\phi_{\mA}\in \mF^1_{\mathfrak{g}}$. Then 
\begin{eqnarray}\label{sum}
Q[\rho_u,\phi_{\mA}]_h=4\phi_{\m[u,\mA]}.
\end{eqnarray}
We know that

\begin{eqnarray}
\m[u,\mA]=(\p_{\mu}\p^{\mu}[u,A_{\nu}]-\p_{\nu}\p_{\mu}[u,A^{\mu}])dx^{\nu}.
\end{eqnarray}
At the same time 

\begin{eqnarray}\label{1}
[Q \rho_u,\phi_{\mA}]_h=2[\phi_{\ud u},\phi_{\mA}]_h=4\psi_{\mY}, 
\end{eqnarray}
where 

\begin{eqnarray}
&&Y_{\nu}=2[\p_{\mu}u, \p^{\mu}A_{\nu}]+2[A_{\mu},\p^{\mu}\p_{\nu}u]+[\p_{\nu}\p_{\mu}u,A^{\mu}]+\nonumber\\
&&[\p_{\nu}A_{\mu},\p^{\mu}u]+[\p_{\mu}\p^{\mu}u,A_{\nu}]+[\p_{\mu}A^{\mu},\p_{\nu}u]
\end{eqnarray}
and

\begin{eqnarray}\label{2}
[\rho_u,Q\phi_{\mA}]_h=4\psi_{[u,\m\mA]}.
\end{eqnarray}
Summing \rf{1} and \rf{2}, we get \rf{sum} and, therefore, the relation \rf{der} also holds in this case. The last 
nontrivial case with $a_1=\rho_u$ is that when $a_2=\psi_{\mW}$. 
We see that 
\begin{eqnarray}
&&Q[\rho_u,\psi_{\mW}]_h=-2\psi_{\di [u,\mW]}=-2\psi_{\ud u\cdot W}+2\psi_{[u,\di\mW]}=\nonumber\\
&&[Q\rho_u,\psi_{\mW}]_h+[\rho_u,Q\psi_{\mW}]_h.
\end{eqnarray}
Let us put $a_1=\phi_{\mA} \in \mF^1_{\mathfrak{g}}$. Then for $a_2=\phi_{\mB} \in \mF^1_{\mathfrak{g}}$, we get 
\begin{eqnarray}
Q[\phi_{\mA}, \phi_{\mB}]_h=-2\chi_{\di\{\mA,\mB\}}.
\end{eqnarray}
We find that 

\begin{eqnarray}
&&\di\{\mA,\mB\}=[\p^{\nu}A_{\mu},\p^{\mu}B_{\nu}]+[A_{\mu},\p^{\mu}\p^{\nu}B_{\nu}+[\p^{\nu}B_{\mu},\p^{\mu}A_{\nu}]+\nonumber\\
&&+[\p_{\nu}\p^{\nu}B_{\mu},A^{\mu}]+[\p_{\nu}B_{\mu},\p^{\nu}A^{\mu}]+\p^{\mu}\p^{\nu}([A_{\mu},B_{\nu}]+[B_{\mu},A_{\nu}]=\nonumber\\
&&[\p^{\nu}\p_{\nu}A_{\mu}-\p_{\mu}\p_{\nu}A^{\nu},B^{\mu}]+[\p^{\nu}\p_{\nu}B_{\mu}-\p_{\mu}\p_{\nu}B^{\nu},A^{\mu}]=\nonumber\\
&&(\m \mA)\cdot \mB+(\m \mB)\cdot \mA.
\end{eqnarray}
This leads to the relation:

\begin{eqnarray}
-2\chi_{\di\{\mA,\mB\}}=-2\chi_{(\m \mA)\cdot \mB}-2\chi_{(\m \mB)\cdot \mA}=[Q\phi_{\mA},\phi_{\mB}]-[\phi_{\mA},Q\phi_{\mB}].
\end{eqnarray}
Therefore, \rf{der} holds in this case. 

It is easy to see that the relation \rf{der}, for the other values of 
$a_1$ and $a_2$,  reduces to a trivial one $0=0$. Thus, we proved \rf{der}.

Let us  switch to the proof of the second relation, including the graded antisymmetric three-linear operation:
\begin{eqnarray}\label{jac}
&&Q[a_1,a_2, a_3]_h+[Q a_1,a_2, a_3]_h+(-1)^{n_{a_1}}[a_1,Q a_2, a_3]_h+\nonumber\\
&&(-1)^{n_{a_1}+n_{a_2}}[ a_1, a_2, Q a_3]_h+[a_1,[a_2, a_3]_h]_h-[[a_1,a_2]_h, a_3]_h-\nonumber\\
&&(-1)^{n_{a_1}n_{a_2}}[a_2,[a_1, a_3]_h]_h=0.
\end{eqnarray}
It is easy to see that \rf{jac} is worth proving in the cases, where $a_1\in \mF^0_{\mathfrak{g}}$, $a_2\in \mF^1_{\mathfrak{g}}$, 
$a_3\in \mF^1_{\mathfrak{g}}$ and $a_1\in \mF^1_{\mathfrak{g}}$, $a_2\in \mF^1_{\mathfrak{g}}$, 
$a_3\in \mF^1_{\mathfrak{g}}$. For the other possible values of $a_1,a_2, a_3$, the relation \rf{jac}
 reduces to the permutations of the above 
two cases or simple consequences of the Jacobi identity for the Lie algebra $\mathfrak{g}$.
 
So, let us consider $a_1=\rho_u\in \mF^0_{\mathfrak{g}}$, $a_2=\phi_{\mA}\in \mF^1_{\mathfrak{g}}$, 
$a_3=\phi_{\mB}\in \mF^1_{\mathfrak{g}}$. In this case, \rf{jac} reduces to
\begin{eqnarray}\label{jac1}
[Q \rho_u,\phi_{\mA}, \phi_{\mB}]_h+[\rho_u,[\phi_{\mA}, \phi_{\mB}]_h]_h-[[\rho_u,\phi_{\mA}]_h,\phi_{\mB} ]_h-
[\phi_{\mA},[\rho_u, \phi_{\mB}]_h]_h=0
\end{eqnarray}
or, rewriting it by means of the expressions for appropriate operations, we get
\begin{eqnarray}
\psi_{\{\ud u,\mA,\mB\}}+\psi_{[u,\{\mA,\mB\}]}-\psi_{\{\mA,[u,\mB]\}}-\psi_{\{\mB,[u,\mA]\}}=0.
\end{eqnarray}
Therefore, to establish \rf{jac1}, one needs to prove that
\begin{eqnarray}
\{\ud u,\mA,\mB\}+[u,\{\mA,\mB\}]-\{\mA,[u,\mB]\}-\{\mB,[u,\mA]\}=0.
\end{eqnarray}
Actually,
\begin{eqnarray}
&&\{\mA,[u,\mB]\}=(2[A_{\mu},[\p^{\mu}u,B_{\nu}]]+2[A_{\mu},[u, \p^{\mu}B_{\nu}]]-\nonumber\\
&&2[\p_{\mu}A_{\nu},[u, B^{\mu}]]+[\p_{\nu}A_{\mu},[u,B^{\mu}]]+[[\p_{\nu}u,B_{\mu}],A^{\mu}]+\nonumber\\
&&[[u,\p_{\nu}B_{\mu}],A^{\mu}]-[A_{\nu},[\p^{\mu}u,B_{\mu}]]-[A_{\nu},[u,\p_{\mu}B^{\mu}]]+\nonumber\\
&&[\p^{\mu}A_{\mu},[u,B_{\nu}]])dx^{\nu}.
\end{eqnarray}
Rearranging the terms and using the Jacobi identity, we find that
\begin{eqnarray}
&&\{\mA,[u,\mB]\}+\{[u,\mB],\mA\}=([u, (2[A_{\mu},\p^{\mu}B_{\nu}]+2[B_{\mu},\p^{\mu}A_{\nu}]+\nonumber\\
&&[\p_{\nu}A_{\mu},B^{\mu}]+[\p_{\nu}B_{\mu},A^{\mu}]-[A_{\nu},\p_{\mu}B^{\mu}]+[\p^{\mu}A_{\mu},B_{\nu}])]+\nonumber\\
&&[A_{\mu},[\p^{\mu}u,B_{\nu}]]+[B_{\mu},[\p^{\mu}u,A_{\nu}]]+
[A_{\mu},[B^{\mu},\p_{\nu}u]]+\nonumber\\
&&[\p^{\mu}u,[B_{\mu},A_{\nu}]]+[B_{\mu},[A^{\mu},\p_{\nu}u]]+[\p^{\mu}u, [A_{\mu},B_{\nu}]])dx^{\nu}=\nonumber\\
&&\{\ud u,\mA,\mB\}+[u,\{\mA,\mB\}].
\end{eqnarray}
In such a way we have  proven \rf{jac1}.

Let us consider the case, where $a_1=\phi_{\mA}\in \mF^1_{\mathfrak{g}}$, $a_2=\phi_{\mB}\in \mF^1_{\mathfrak{g}}$, 
$a_3=\phi_{\mC}\in \mF^1_{\mathfrak{g}}$. For this choice of variables, \rf{jac} has the  form
\begin{eqnarray}\label{jac2}
&&Q[\phi_{\mA},\phi_{\mB}, \phi_{\mC} ]_h+[\phi_{\mA},[\phi_{\mB}, \phi_{\mC}]_h]_h-[[\phi_{\mA},\phi_{\mB}]_h, \phi_{\mC}]_h+\nonumber\\
&&[\phi_{\mB},[\phi_{\mA},\phi_{\mC} ]_h]_h=0
\end{eqnarray}
or, on the level of differential forms,
\begin{eqnarray}\label{formjac}
\di\{\mA,\mB,\mC\}+\mA\cdot \{\mB,\mC\}+ \mC\cdot\{\mA,\mB\}+\mB\cdot \{\mC,\mA\}=0.
\end{eqnarray}
To prove \rf{formjac}, we write the expression for $\mC\cdot\{\mA,\mB\}$:
\begin{eqnarray}\label{abc}
&&\mC\cdot\{\mA,\mB\}=2[C^{\nu},[A_{\mu},\p^{\mu}B_{\nu}]]-2[C^{\nu},[\p_{\mu}A_{\nu},B^{\mu}]]+\nonumber\\
&&2[C^{\nu},[\p_{\nu}A^{\mu},B_{\mu}]]+[C^{\nu},[\p_{\nu}B_{\mu},A^{\mu}]]-[C^{\nu},[A_{\nu},\p^{\mu}B_{\mu}]]+\nonumber\\
&&[C^{\nu},[\p^{\mu}A_{\mu},B_{\nu}]]=-([C^{\nu},[\p^{\mu}B_{\nu},A_{\mu}]]+[C^{\nu},[\p_{\mu}A_{\nu},B^{\mu}]]+\nonumber\\
&&[\p^{\mu}B_{\nu},[C^{\nu},A_{\mu}]]+[\p_{\mu}A_{\nu},[C^{\nu},B^{\mu}]]+[C^{\nu},[A_{\nu},\p_{\mu}B^{\mu}]]+\nonumber\\
&&[C^{\nu},[B_{\nu},\p_{\mu}A^{\mu}]])+[A_{\mu},[C^{\nu},\p^{\mu}B_{\nu}]]+[B_{\mu},[C^{\nu},\p_{\mu}A_{\nu}]]-\nonumber\\
&&[C^{\mu},[A_{\nu},\p_{\mu}B^{\nu}]]-[C^{\mu},[B_{\nu},\p_{\mu}A^{\nu}]].
\end{eqnarray}
In order to obtain the last equality, we have used the Jacobi identity from $\mathfrak{g}$. Now, we observe that adding to \rf{abc} its 
cyclic permutations, that is $\mA\cdot \{\mB,\mC\}$ and $\mB\cdot \{\mA,\mC\}$, we find that the sum of cyclic permutations 
of terms in round brackets 
(see the last equality of \rf{abc}) gives $\di \{\mA,\mB,\mC\}$ while all other terms cancel. This proves  
the relation \rf{formjac} and, 
therefore, 
\rf{jac2}. Hence we proved \rf{jac}.  

The relations left are
\begin{eqnarray}
&&[b,[a_1,a_2, a_3]_h]_h-(-1)^{n_b(n_{a_1}+n_{a_2}+n_{a_3})}[a_1,[a_2, a_3, b]_h]_h+\nonumber\\
&&(-1)^{n_{a_2}(n_{b}+n_{a_1})}[a_2,[b,a_1, a_3]_h]_h-(-1)^{n_{a_3}(n_{a_1}+n_{a_2}+n_{b})}
[a_3,[b, a_1,a_2]_h]_h\nonumber\\
&&=[[b,a_1]_h,a_2, a_3]_h+(-1)^{n_{a_1}n_{b}}[a_1,[b,a_2]_h, a_3]_h+\nonumber\\
&&(-1)^{(n_{a_1}+n_{a_2})n_{b}}[a_1,a_2, [b,a_3]_h]_h,\nonumber\\
&&[[a_1,a_2, a_3]_h,b,c]_h=0.
\end{eqnarray}
However, to prove the first one, it is easy to see that this 
relation is nontrivial only, when $b\in \mF^0_{\mathfrak{g}}$ or 
$b\in \mF^1_{\mathfrak{g}}$and $a_i\in\mF^1_{\mathfrak{g}}$. 
Therefore, it becomes a consequence of the Jacobi identity from $\mathfrak{g}$. The second one 
is trivial since the three-linear operation 
takes values in $\mF^2_{\mathfrak{g}}$ and it is zero for any argument lying in $\mF^2_{\mathfrak{g}}$. 
\break Proposition 2.3 is proven.
\end{proof}

\vspace{3mm}

\noindent {\bf Proposition A.2.} {\it The multilinear products introduced in Definition 2.5 are graded as antisymmetric, i.e.}
\begin{eqnarray}\label{sym}
\{a_1,...,a_i,a_{i+1},..., a_n\}_h=-(-1)^{n_{a_i}n_{a_{i+1}}}\{a_1,...,a_{i+1},a_{i},..., a_n\}_h,
\end{eqnarray}
{\it where $n_{a_i}$ denotes the ghost number of $a_i$.}
\begin{proof}
 For the bilinear form $\{\cdot, \cdot\}$, the statement is the direct consequence of Proposition 2.2. 
To prove \rf{sym} of the three-linear form $\{\cdot, \cdot, \cdot\}$, we need to check four relations:
\begin{eqnarray}
&&\langle[\rho_u, \rho_v]_h,\chi_a\rangle=\langle[\chi_a, \rho_u]_h,\rho_v\rangle, \quad 
\langle[\phi_{\mA}, \phi_{\mB}]_h,\phi_{\mC}\rangle=\langle[\phi_{\mA}, \phi_{\mC}]_h,\phi_{\mB}\rangle,\nonumber\\
&&\langle[\rho_u, \phi_{\mA}]_h,\psi_{\mW}\rangle= \langle[\psi_{\mW},\rho_u]_h,\phi_{\mA}\rangle=
\langle[\phi_{\mA},\psi_{\mW}]_h,\rho_{u}\rangle.
\end{eqnarray}
Almost all of them are the simple consequence of the basic property of the invariant form: 
\begin{eqnarray}\label{kil}
(X,[Y,Z])_K=(Z,[X,Y])_K,
\end{eqnarray}
where $X,Y,Z \in \mg$. 
The only nontrivial one is 
\begin{eqnarray}
\langle[\phi_{\mA}, \phi_{\mB}]_h,\phi_{\mC}\rangle=\langle[\phi_{\mA}, \phi_{\mC}]_h,\phi_{\mB}\rangle.
\end{eqnarray}
Let us prove it. From the definition of the inner product and bilinear operation in the homotopy Lie superalgebra 
we see, that it is equivalent to the following statement:
\begin{eqnarray}
\int d^Dx(\{\mA,\mB\},\mC)(x)=\int d^Dx(\{\mA,\mC\},\mB)(x)
\end{eqnarray}
Let us write it explicitly:
\begin{eqnarray}
&&\int d^Dx(\{\mA,\mB\},\mC)(x)=\int d^Dx (C^{\nu}(x), 2[A_{\mu},\p^{\mu}B_{\nu}](x)+2[B_{\mu},\p^{\mu}A_{\nu}](x)\nonumber\\
&&+[\p_{\nu}A_{\mu},B^{\mu}](x)
+[\p_{\nu}B^{\mu},A_{\mu}](x)+[\p^{\mu}A_{\mu},B_{\nu}](x)+[\p^{\mu}B_{\mu},A_{\nu}](x))_K=\nonumber\\
&&\int d^Dx (B^{\nu}(x),\p^{\mu}[A_{\mu},C_{\nu}](x)-2[C^{\mu},\p_{\nu}A_{\mu}](x)-[\p_{\mu}A_{\nu},C^{\mu}](x)+\nonumber\\
&&\p^{\mu}[C_{\mu},A_{\nu}](x)+[C_{\nu},\p_{\mu}A^{\mu}](x)+\p^{\nu}[C^{\mu},A_{\mu}](x))_K=\nonumber\\
&&\int d^Dx (B_{\nu}(x),2[A_{\mu},\p^{\mu}C^{\nu}](x)+2[C_{\mu},\p^{\mu}A^{\nu}](x)+[\p^{\mu}A_{\mu},C^{\nu}](x)+\nonumber\\
&&[\p^{\mu}C_{\mu},A^{\nu}](x)+[\p^{\nu}A_{\mu},C^{\mu}](x)+[\p^{\nu}C_{\mu},A^{\mu}](x))_K=\nonumber\\
&&\int d^Dx(\{\mA,\mC\},\mB)(x).
\end{eqnarray}
Thus, we have proven the proposition for three-linear form. So, to finish the proof, we need to check the relation \rf{sym} for the 
4-linear product. In other words, we need to show that 
\begin{eqnarray}
\{\phi_{\mA},\phi_{\mB},\phi_{\mC},\phi_{\mD}\}=\{\phi_{\mA},\phi_{\mB},\phi_{\mD},\phi_{\mC}\}.
\end{eqnarray}
By definition, this is equivalent to the relation
\begin{eqnarray}
\int d^D x(\{\mA,\mB,\mC\},\mD)(x)=\int d^D x(\{\mA,\mB,\mD\},\mC)(x),
\end{eqnarray}
which can easily be shown to be true by the iterated use of \rf{kil}. 
\break Thus, Proposition 2.4 is proven. 
\end{proof}

\end{document}